\documentstyle[12pt,times,epsf,psfig]{article}

\begin{document}

\baselineskip 6mm
\renewcommand{\thefootnote}{\fnsymbol{footnote}}

%------------ Sangmin's macro's, etc  -----------

\newcommand{\nc}{\newcommand}
\newcommand{\rnc}{\renewcommand}

%\headheight=0truein
%\headsep=0truein
%\topmargin=0truein
%\oddsidemargin=0truein
%\evensidemargin=0truein
%\textheight=9truein
%\textwidth=6.5truein

\rnc{\baselinestretch}{1.24}	% 1.5 spacing btwn text lines
\setlength{\jot}{6pt} 		% spacing btwn the rows of an eqnarray
\rnc{\arraystretch}{1.24}   	% spacing btwn the rows of a non-eqn array

%%%%%%%%%%%%%%%%%%%%%% Equation Numbering %%%%%%%%%%%%%%%%%%%%%%%
\makeatletter
\rnc{\theequation}{\thesection.\arabic{equation}}
\@addtoreset{equation}{section}
\makeatother                      

%%%%%%%%%%%%%%%%%%%%%%%%%%%%%%%%%%%%%%%%%%%%%%%%%%%%%%%%%%%%%%%%%
%								%
%		NEW COMMANDS AND MACROS				%
%								%
%%%%%%%%%%%%%%%%%%%%%%%%%%%%%%%%%%%%%%%%%%%%%%%%%%%%%%%%%%%%%%%%%

%%%%% Simplify some frequently used LaTeX commands %%%%%

\nc{\be}{\begin{equation}}
\nc{\ee}{\end{equation}}
\nc{\bea}{\begin{eqnarray}}
\nc{\eea}{\end{eqnarray}}
\nc{\xx}{\nonumber\\}

\nc{\eq}[1]{(\ref{#1})}
%\nc{\newcaption}[1]{\centerline{\parbox{6in}{\caption{#1}}}}

%\nc{\fig}[3]{
%\begin{figure}
%\centerline{\epsfxsize=#1\epsfbox{#2.eps}}
%\newcaption{#3. \label{#2}}
%\end{figure}
%}

%%%%% Journal Macros %%%%%

\nc{\np}[3]{Nucl. Phys. {\bf B#1} (#2) #3}
\nc{\pl}[3]{Phys. Lett. {\bf #1B} (#2) #3}
\nc{\prl}[3]{Phys. Rev. Lett.{\bf #1} (#2) #3}
\nc{\prd}[3]{Phys. Rev. {\bf D#1} (#2) #3}
\nc{\ap}[3]{Ann. Phys. {\bf #1} (#2) #3}
\nc{\prep}[3]{Phys. Rep. {\bf #1} (#2) #3}
\nc{\rmp}[3]{Rev. Mod. Phys. {\bf #1} (#2) #3}
\nc{\cmp}[3]{Comm. Math. Phys. {\bf #1} (#2) #3}
\nc{\mpl}[3]{Mod. Phys. Lett. {\bf #1} (#2) #3}
\nc{\cqg}[3]{Class. Quant. Grav. {\bf #1} (#2) #3}
\nc{\jhep}[3]{J. High Energy Phys. {\bf #1} (#2) #3}

%%%%% Special Letters

\def\vare{\varepsilon}
\def\bz{\bar{z}}
\def\bw{\bar{w}}

%%% Caligraphic letters %%%%

\def\CA{{\cal A}}
\def\CC{{\cal C}}
\def\CD{{\cal D}}
\def\CE{{\cal E}}
\def\CF{{\cal F}}
\def\CG{{\cal G}}
\def\CI{{\cal I}}
\def\CT{{\cal T}}
\def\CM{{\cal M}}
\def\CN{{\cal N}}
\def\CP{{\cal P}}
\def\CL{{\cal L}}
\def\CV{{\cal V}}
\def\CS{{\cal S}}
\def\CW{{\cal W}}
\def\CY{{\cal Y}}
\def\CS{{\cal S}}
\def\CO{{\cal O}}
\def\CP{{\cal P}}
\def\CN{{\cal N}}

%%% Double line letters %%%

\def\IR{{\hbox{{\rm I}\kern-.2em\hbox{\rm R}}}}
\def\IB{{\hbox{{\rm I}\kern-.2em\hbox{\rm B}}}}
\def\IN{{\hbox{{\rm I}\kern-.2em\hbox{\rm N}}}}
\def\IC{\,\,{\hbox{{\rm I}\kern-.59em\hbox{\bf C}}}}
\def\IZ{{\hbox{{\rm Z}\kern-.4em\hbox{\rm Z}}}}
\def\IP{{\hbox{{\rm I}\kern-.2em\hbox{\rm P}}}}
\def\IH{{\hbox{{\rm I}\kern-.4em\hbox{\rm H}}}}
\def\ID{{\hbox{{\rm I}\kern-.2em\hbox{\rm D}}}}

%%% Greek letters %%%

\def\a{\alpha}
\def\b{\beta}
\def\ga{\gamma}
\def\d{\delta}
\def\ep{\epsilon}
\def\ph{\phi}
\def\k{\kappa}
\def\l{\lambda}
\def\m{\mu}
\def\n{\nu}
\def\th{\theta}
\def\rh{\rho}
\def\s{\sigma}
\def\t{\tau}
\def\w{\omega}
\def\G{\Gamma}

%%%%% Mathematical Symbols

\def\half{\frac{1}{2}}
\def\imp{\Longrightarrow}
\def\dint#1#2{\int\limits_{#1}^{#2}}
\def\goto{\rightarrow}
\def\para{\parallel}
\def\brac#1{\langle #1 \rangle}
\def\del{\nabla}
\def\grad{\nabla}
\def\curl{\nabla\times}
\def\div{\nabla\cdot}
\def\p{\partial}
\def\e{\epsilon_0}

%%%%% Roman font in math

\def\Tr{{\rm Tr}}
\def\det{{\rm det}}
\def\im{{\rm Im~}}
\def\re{{\rm Re~}}

%%%%% Names

\def\Kahler{K\"{a}hler}

%%%%% For this paper only

\def\e{\varepsilon}
\def\bA{\bar{A}}
\def\c{\zeta}

\begin{titlepage}
%---------------- preprint number ---------------
\hfill\parbox{4cm}
{hep-th/0011019}

%------------------------ title ------------------------
\vspace{15mm}
\centerline{\Large \bf Stretched strings and worldsheets
with a handle}
%---------------- authors and addresses ----------------
\vspace{10mm}
\begin{center}  
Youngjai Kiem\footnote{ykiem@newton.skku.ac.kr},
Dong Hyun Park\footnote{donghyun@newton.skku.ac.kr},
and Haru-Tada Sato\footnote{haru@taegeug.skku.ac.kr}\\[2mm] 
{\sl BK21 Physics Research Division and Institute of Basic Science, 
Sungkyunkwan University, 
Suwon 440-746, Korea}
\end{center}
\thispagestyle{empty}
\vskip 40mm

%----------------------- abstract ----------------------
\centerline{\bf ABSTRACT}
\vskip 5mm
\noindent
In the presence of the constant background NS two-form
gauge field, we construct the worldsheet partition functions,
bulk propagators and boundary propagators for the worldsheets
with a handle and a boundary.  We analyze the noncommutative 
$\phi^3$ field theory amplitudes that correspond to the 
general two-point insertions on the two-loop nonplanar
vacuum bubble.  By the direct string theory amplitude 
computations on the worldsheets with a handle, which
reduce to the aforementioned field theory amplitudes in
the decoupling limit, we find that the stretched string
interpretation remains valid for the types of amplitudes in
consideration.  This completes the demonstration that the 
stretched string picture holds up in the general multiloop
context. 
\vspace{2cm}
\end{titlepage}

%-------------------------------------------------------
\baselineskip 7mm
\renewcommand{\thefootnote}{\arabic{footnote}}
\setcounter{footnote}{0}

%%%%%%%%%%%%%%%%%%%%%%%%%%%%%%%%%%%%%%%%%%%%%%%%%%%%%%%%%%%%%%
\section{Introduction}
%%%%%%%%%%%%%%%%%%%%%%%%%%%%%%%%%%%%%%%%%%%%%%%%%%%%%%%%%%%%%%

The noncommutative field theories resulting from a certain
decoupling limit of the open string theory with the constant 
background NS two-form gauge field \cite{string,witten} have 
an inherent nonlocality \cite{nonlocal}.
Such unconventional features as the UV/IR mixing in 
noncommutative field theories \cite{uvir} have largely been 
attributed to it.  From the underlying string theory point of view, 
the stretched string interpretation of \cite{liu} has been successfully 
applied to explain that character of noncommutative field theories.  
In particular, in \cite{klp}, the original suggestion of \cite{liu} 
based on the one-loop analysis \cite{oneloop} was extended to 
the multiloop context 
for the amplitudes coming from the (non)planar external vertex 
insertions on planar vacuum diagrams.  The main theme of this note 
is to extend the analysis to the case of the external insertions 
on nonplanar vacuum diagrams.  We find that the stretched string 
interpretation can successfully be extended to the case in
consideration.

The technical highlight of this note is the explicit construction 
of the worldsheet partition function and the propagators for the
open string worldsheets with a handle attached, presented in Section 2.
Our construction directly computes the (boundary) open string 
propagators as well as the (bulk) closed string propagators.  
In the Reggeon vertex formalism of \cite{chu}, one
computes the amplitudes and the propagators are read off from those
expressions\footnote{In \cite{chu}, only open string insertions 
were analyzed.  However, by using the techniques developed in, for example, 
\cite{chu1} in the commutative context, one may consider the closed 
string vertex insertions in the noncommutative context as well.  
We thank R.~Russo for pointing this out to us.}.
In section 3, we present a field theory analysis covering all the 
two-point 1PI external insertions on the two-loop nonplanar vacuum 
bubble in the noncommutative $\phi^3$ theory.  Armed with the 
results in section 2, 
we then compute the string theory amplitudes involving nonplanar 
worldsheets, and consider the field theory reduction of the 
string theory calculations; we demonstrate the validity of the 
stretched string interpretation for the amplitudes in consideration
in the following sense. 
Typical two-loop vacuum bubbles in the noncommutative $\phi^3$ field
theory are depicted in Fig.~1, a planar vacuum bubble and 
a nonplanar vacuum bubble.  When extended to string theory diagrams
by `thickening' the Feynman diagrams, a nonplanar vacuum bubble
corresponds to an open string worldsheet with a handle 
attached.  As shown in Fig.~1, the nonplanar vacuum bubble 
can also be regarded as coming from the nonplanar one-loop amplitude
with the external vertices connected.  In this sense, the
one-loop external momentum turns into an internal momentum that
should be integrated over all values.  Since the stretching 
of the open string is given by 
$\Delta X^{\mu} = \theta^{\mu \nu} p_{\nu}$, in the decoupling
limit $\alpha^{\prime} \rightarrow 0$, the stretching length 
$\Delta X^{\mu} G_{\mu \nu} \Delta X^{\nu}$ (here $G_{\mu \nu}$ is
the open string metric) for the external open string can be chosen 
to be larger than the string scale $\alpha^{\prime}$.  However,
as a loop momentum, the contribution to the amplitudes
from the $  \Delta X^{\mu} G_{\mu \nu} \Delta X^{\nu}
 < \alpha^{\prime}$ momentum regime should also be considered.
What we show in section 3 is that this type of contribution
always goes to zero in the decoupling limit $\alpha^{\prime}
\rightarrow 0$.  We note that the field theory results
presented in section 3 are consistent with those of \cite{roiban}.
In fact, our string theory consideration shows that the analysis of 
\cite{roiban} is natural from the underlying string theory
point of view.   

In section 4, we discuss our results and the possible applications.

\begin{figure}\label{fig1}
\centerline{\psfig{file=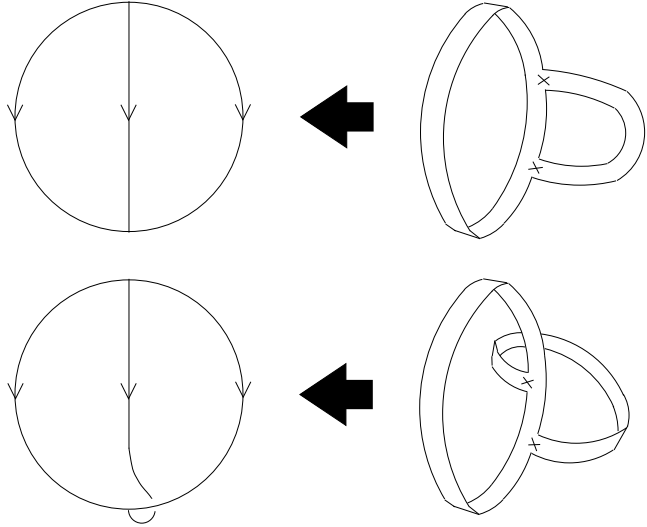,height=7cm}}
\caption{Two two-loop vacuum bubbles in $\phi^3$ field
theory are shown.  The upper diagram is a planar vacuum
bubble and the lower diagram is a nonplanar vacuum bubble.
The `thickened' version of the diagrams are also shown.}
\end{figure}

%%%%%%%%%%%%%%%%%%%%%%%%%%%%%%%%%%%%%%%%%%%%%%%%%%%%%%%%%%%%%%
\section{Worldsheet partition function and propagators}
%%%%%%%%%%%%%%%%%%%%%%%%%%%%%%%%%%%%%%%%%%%%%%%%%%%%%%%%%%%%%%

In this section, after reviewing the geometries of worldsheets
with and without a handle, we construct the worldsheet partition 
functions and propagators.  For the latter, appropriate forms 
for both
bulk and boundary propagators are computed.  Though the essential
part of our analysis can be generalized to worldsheets with 
multiple handles, we restrict our attention only to the
worldsheets with a single handle and a boundary, where a simple
and explicit analysis is possible.   

%%%%%%%%%%%%%%%%%%%%%%%%%%%%%%%%%%%%%%%%%%%%%%%%%%%%%%%%%%%%%%
\subsection{$g=b=1$ Worldsheets and partition functions}
%%%%%%%%%%%%%%%%%%%%%%%%%%%%%%%%%%%%%%%%%%%%%%%%%%%%%%%%%%%%%%

A useful way to construct an open string worldsheet is
to start from a closed string worldsheet and to fold it
by half.  From here on, we denote the genus $g$ worldsheet
with $b$ boundaries as $(gb)$ surface.  In Fig.~2, 
one finds a schematic representation of the closed string 
$(20)$ worldsheet.  As denoted in the figure, the canonical
basis of homology cycles is given by $a_{\alpha}$ and
$b_{\alpha}$ cycles where $\alpha = 1$ and $2$, with the
intersection parings
\begin{equation}
(a_{\alpha} , a_{\beta} ) = (b_{\alpha} , b_{\beta} ) ~~ , ~~
(a_{\alpha} , b_{\beta} ) = - (b_{\alpha} , a_{\beta} )
= \delta_{\alpha \beta} ~~ \rightarrow 
J = \pmatrix{ 0_2 & 1_2 \cr -1_2 & 0_2 } ~ .
\end{equation}
Here $0_2$ and $1_2$ denote the $2 \times 2$ 
zero and identity matrices, respectively.  
Dual to these cycles, there exist two holomorphic 
(antiholomorphic) one-forms $\omega_{\alpha}$ 
($\bar{\omega}_{\alpha}$) among the cohomology group 
elements.  The period matrix $\tau$ and the normalization 
of these one-forms are given by
\begin{equation}
 \oint_{a_{\alpha}} \omega_{\beta} = \delta_{\alpha \beta} ~~~ , ~~~
 \oint_{b_{\alpha}} \omega_{\beta} = \tau_{\alpha \beta} ~ .
\end{equation}
Up to three loops, it is known that the moduli space of
the worldsheets are parameterized by the symmetric 
period matrix without any redundancy.  

\begin{figure}\label{torus}
\centerline{\psfig{file=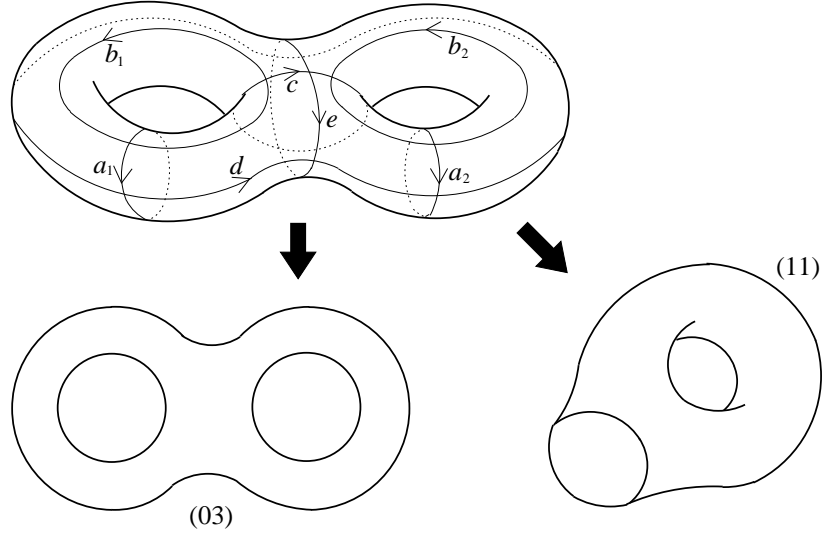,height=7cm}}
\caption{Depending on how we fold a $(20)$ surface,
either a $(03)$ surface or a $(11)$ surface is obtained.
The various homology cycles of the $(20)$ surface
are also depicted.}
\end{figure}

Two inequivalent open string worldsheets that can be
obtained from $(20)$ worldsheet by folding are $(03)$
surface and $(11)$ surface, each corresponding to a
planar two-loop vacuum bubble and a nonplanar two-loop
vacuum bubble.  To be precise, the folding operation
is an orientation-reversing (anticonformal) involution
map $I$ ($I^2 = {\rm identity} $) and we identify the
point $p$ with its involution image $I(p)$.  The fixed points 
under the 
involution correspond to worldsheet boundaries.
When acting on homology cycles, $I$ is represented by
a matrix
\begin{equation}
 \pmatrix{ a^{\prime} \cr b^{\prime} } = I
 \pmatrix{ a \cr b } = \pmatrix{ H & G \cr F & E }
 \pmatrix{ a \cr b } 
\label{Imat}
\end{equation}
and it satisfies 
\begin{equation}
  IJI^T = - J ~ .
\end{equation}
Among the period matrix elements, only the ``even" sector under 
the involution $I$ survives the folding, namely,
\begin{equation}
 \tau = ( E \bar{\tau} + F ) ( G \bar{\tau} + H )^{-1} ,
\end{equation}
which reduces the real six dimensional moduli space of
$(20)$ surfaces to the real three dimensional moduli
spaces of $(11)$ surfaces or $(03)$ surfaces.  

>From what follows, we will concentrate on $(11)$ surfaces.
Therefore, when acting on the canonical homology cycles,
the involution matrix $I$ can be written as
\begin{equation}
 \pmatrix{ a^{\prime} \cr b^{\prime} }
 = I \pmatrix{ a \cr b }
 = \pmatrix{ \sigma & 0_2 \cr 0_2 & - \sigma }
   \pmatrix{ a \cr b  } ~~~ , ~~~ \sigma \equiv 
  \pmatrix{ 0 & 1 \cr 1 & 0} ~~~ , 
\label{involpr}
\end{equation}
which yields the period matrix of the form
\begin{equation}
 \tau = \pmatrix{ a + ib & ic \cr ic & -a + i b } ,
\end{equation}
where $a$, $b$, $c$ are real numbers.  Even if this
basis is easier to visualize, for the further analysis, we 
find it much simpler to 
use a different basis for the homology cycles.  With
an $Sp(4, \IZ)$ matrix $M$ (satisfying $MJM^T = J$ 
and thus preserving
the intersection pairing), we change the basis into
\begin{equation}
 \pmatrix{ \tilde{a} \cr \tilde{b}  }
 = M  \pmatrix{ a \cr  b }
 = \pmatrix{ D & C \cr B & A }  \pmatrix{ a \cr b  }
 = \pmatrix{ 1 & -1 & 0 & 0 \cr 0 & 0 & 1 & 1 \cr
             0 & 0 & 1 & 0 \cr 0 & -1 & 0 & 0 }
   \pmatrix{ a_1 \cr a_2 \cr 
           b_1 \cr b_2 } ~ .
\end{equation}
When normalized in the new basis, the period matrix
$\tilde{\tau}$ ($\tilde{\tau} = ( A \tau + B )
( C \tau + D )^{-1}$) and the involution matrix $\tilde{I}$ 
($\tilde{I} = M I M^{-1}$) can be written as
\begin{equation}
\tau = \pmatrix{ i T_{11}  & \frac{1}{2} + i T_{12} \cr
                 \frac{1}{2} + iT_{21} & i T_{22} } 
\label{period}
\end{equation}
and
\begin{equation}
  I = \pmatrix{ -1_2 & 0_2 \cr - \sigma & 1_2 } ~ ,
\label{invol}
\end{equation}
where $T_{11}$, $T_{22}$ and
$T_{12} = T_{21}$ are three real numbers.  In (\ref{period})
and (\ref{invol}), since we will stick to the new basis hereafter, 
we have dropped tildes denoting the cycles, period matrix, etc.~, 
in this basis.  When compared to the $(03)$ surfaces (in the 
canonical homology basis) where 
\begin{equation}
\tau = \pmatrix{ i T_{11}  &  i T_{12} \cr
                 i T_{21} & i T_{22} }  ~~~ , ~~~
  I = \pmatrix{ -1_2 & 0_2 \cr  0_2 & 1_2 } ~ ,
\label{03tau}
\end{equation}
we immediately note that the $(11)$ surfaces come from
the nonplanar two-point (open string) vertex insertions
on an annulus (worldsheet vertex separation given by $\Delta z
=  1/2 + i T_{12}$), while the $(03)$ surfaces 
originate from the planar two-point (open string) vertex
insertions with the vertex separation $\Delta z = i T_{12}$.

We first consider the case when there is no background
NS two-form field.  Our situation of interests is the setup
where there are $N$ stack of parallel D$p$-branes.  
It is then known from Refs.~\cite{blau,bianchi} 
that the partition function for $(11)$ surfaces with
the period matrix (\ref{period}) can be written
as (up to an overall normalization factor)
\begin{equation}
Z_{(11)} = N \int d T_{11} d T_{22} d T_{12}
 \frac{ |W( \tau )| }{(\det ~ 2 \pi 
 \alpha^{\prime} ~ \im \tau )^{(p+1)/2}} ~,
\label{0part}
\end{equation}
where 
\[  | W( \tau ) | = \prod_{a=1}^{10} 
    | \theta_a ( 0 | \tau ) |^{-2}  \]  
and $\theta_a$'s are the ten even Riemann theta functions for the
(20) surfaces.  Similarly the partition function for
$(03)$ surfaces can be written in the same form as (\ref{0part}) 
with the period matrix (\ref{03tau}).  We note that the partition 
function (\ref{0part}) is valid only when the involution matrix 
$I$ in (\ref{Imat}) has components $G = 0$ and $H = - E = - 1_2$.  
Clearly, both the involution matrices of (\ref{invol}) and 
(\ref{03tau}) satisfy this requirement, unlike the case 
of (\ref{involpr}).  

The key issue is to study the modification of the partition
function when we turn on the constant background NS two-form
field ($B$).  As was argued in Refs.~\cite{klp,andreev}, the 
partition functions for the {\em planar} ($g=0$) worldsheets do
not change at all (modulo the overall multiplication factor)
as we turn on the $B$ field.  However for the {\em nonplanar}
worldsheets such as (11) surfaces, where there exist intersecting
cycles, there are changes in the
form of the partition function \cite{chu}; 
we instead have the following expression
\begin{equation}
Z_{(11)} = N \int d T_{11} d T_{22} d T_{12}
 \frac{ |W( \tau )| }{ \sqrt{ \det ~ ( 2 \pi \alpha^{\prime} ~ 
  G_{\mu \nu} ~ \im \tau 
 + \frac{i}{2} \theta_{\mu \nu} \CI )} } ~ .
\label{parti}
\end{equation}
The open string metric $G^{\mu \nu}$ and the noncommutativity
parameter $\theta^{\mu \nu}$ are related to the corresponding
closed string quantities via
\begin{equation}
 G^{\mu \nu} = (g_{\mu \nu} + B_{\mu \nu})^{-1}_S  ~~~ , ~~~
 \theta^{\mu \nu} =  2 \pi \alpha^{\prime} ( g_{\mu \nu} +
  B_{\mu \nu} )^{-1}_A ~,
\end{equation}
where the subscripts $S$ and $A$ denote the symmetric and
the antisymmetric parts of a matrix, respectively.  The 
$ 2 \times 2$ matrix $\CI$ is the intersection matrix for 
the intersecting cycles
that are present in the worldsheet
\begin{equation}
 \CI = \pmatrix{ 0 & 1 \cr -1 & 0} ~ .
\label{interm}
\end{equation}
In (\ref{parti}), the determinant is taken with respect to the
$2(p+1) \times 2(p+1)$ matrix $2 \pi \alpha^{\prime} ~ 
 G_{\mu \nu} ~ \im \tau 
 + \frac{i}{2} \theta_{\mu \nu} \CI$; as such, when 
$B_{\mu \nu} = 0$, the partition function (\ref{parti})
reduces to (\ref{0part}).  As is clear from the mode expansion 
at the tree level \cite{string} 
and one-loop worldsheet propagators \cite{oneloop}, the zero mode 
parts of the string modes are what is affected by turning on
the $B$ field.  Furthermore, the knowledge of one-loop worldsheet 
propagator is enough to see that the two-loop partition function 
(\ref{parti}) is the correct one, as sketched in Appendix A. 

%%%%%%%%%%%%%%%%%%%%%%%%%%%%%%%%%%%%%%%%%%%%%%%%%%%%%%%%%%%%%%
\subsection{Worldsheet propagators}
%%%%%%%%%%%%%%%%%%%%%%%%%%%%%%%%%%%%%%%%%%%%%%%%%%%%%%%%%%%%%%

The knowledge of the worldsheet partition function is
helpful for the construction of the worldsheet propagators.
We suppose that the $\theta_{\mu \nu}$ matrix is
$2 \times 2$ block-diagonalized by an appropriate choice of the
target space coordinates.  Then for each block with
$\theta_{\mu ~ \mu+1} = \theta_{\mu}$ (for odd $\mu$), we can compute
the inverse of the matrix  $ 2 \pi \alpha^{\prime} ~ 
 G_{\mu \nu} ~ \im \tau 
 + \frac{i}{2} \theta_{\mu \nu} \CI$ involved in the partition 
function (\ref{parti}):
\begin{equation}
 \pmatrix{ 2 \pi \alpha^{\prime} \im \tau & 
  \frac{i}{2} \theta_{\mu} \CI \cr 
      - \frac{i}{2} \theta_{\mu} \CI & 2 \pi \alpha^{\prime} ~
   \im \tau }^{-1}
 = \pmatrix{ \tilde{T}_{\theta_{\mu}}^{-1}  &  
    - \frac{i}{2 D_{\theta_{\mu}} }
 \theta_{\mu} \CI \cr  \frac{i}{2 D_{\theta_{\mu}}}
  \theta_{\mu} \CI  &
   \tilde{T}_{\theta_{\mu}}^{-1} },
\label{useful}
\end{equation}
where we introduce
\begin{equation}
 D_{\theta_{\mu}} = (2 \pi \alpha^{\prime} )^2 ( T_{11} T_{22} - T_{12}^2 )
   + \frac{1}{4} \theta_{\mu}^2 ~~~ , ~~~
  \tilde{T}_{\theta_{\mu}}^{-1} = \frac{2 \pi \alpha^{\prime}}
  {D_{\theta_{\mu}}} 
  \pmatrix{T_{22} & - T_{12} \cr -T_{12} & T_{11} } ~ .
\label{deff}
\end{equation}
We note that the matrix in (\ref{useful}) is a matrix in
the target space coordinate basis, while the basis of
the matrix in (\ref{deff}) is the homology cycle basis. 

For simplicity, we start our consideration from the case
when the only nonzero component of the $B$-field is
$B_{12} = B$.  Furthermore, we suppose $2 \pi \alpha^{\prime} =1$,
the closed string metric $g_{\mu \nu} = \eta_{\mu \nu}$ and
the open string metric is given by $G^{\mu \nu} = \eta^{\mu \nu}
/ (1 +B^2 )$, which also implies that $\theta_{\mu}^2 
 \equiv \theta^{12} \theta_{12} = B^2$.  Under these conventions,
we note that $D_B =  T_{11} T_{22} - T_{12}^2 + B^2 /4$ and
$D_0 =  T_{11} T_{22} - T_{12}^2$.  
For the $(03)$ surfaces, the propagators
for $X = X^1$ and $Y= X^2$ are given by \cite{klp}
\begin{eqnarray}
\label{xxprop}
\langle X (z) X (z') \rangle \!\!\! &=& \!\!\! G(z, z') \! + \!
\frac{1 \! - \! B^2}{1 \! + \! B^2} G(z, \bar{z'})  \! + \!
\frac{2  B^2}{1 \! + \! B^2}  \re \Omega_\a 
 (\tilde{T}^{-1}_0 )^{\alpha \beta} \re \Omega_\b ,
\\
\label{xyprop}
\langle Y (z) X (z') \rangle \!\!\! &=& \!\!\!
\frac{2B}{1 \! + \! B^2}
\left( \frac{1}{2 \pi} \log \frac{E(z, \bar{z'})}
  { \left( E ( z, \bar{z'} ) \right)^* }
 + 2 i  \re \Omega_\a (\tilde{T}^{-1}_0 )^{\alpha \beta}  
\im \Omega_\b ~ \right)~,
\end{eqnarray}
where the function $G$ is defined as
\begin{equation}
G(z,z') =  - \frac{1}{2\pi} \log \left| E (z, z') \right|^2 + 
 \im \Omega_\a (\tilde{T}^{-1}_0 )^{\a\b} \im \Omega_\b ~.
\end{equation}
The overbar on the worldsheet position denotes the involution 
transformed position $\bar{z} = I(z)$ of the position $z$.  
The indices $(\a , \b )$ run over the $(1,2)$ homology
cycles, $E(z , w | \tau )$ is the prime form on $(20)$ surface,
and $\Omega_{\alpha}$ is the complex integral of the
Abelian differential $\omega_{\alpha}$ from a point $\bar{z}^{\prime}$
to a point $z$ along a contractible path
\begin{equation}
 \Omega_{\a} = \int_{\bar{z}^{\prime}}^{z} \omega_\a ~ ,
\end{equation} 
where the path passes through a reference point $P$ lying on
one of the boundaries.  We note that for a contractible path
\begin{equation}
  \im \int_{z^{\prime}}^z \omega_{\alpha} 
 =  \im \int_{\bar{z}^{\prime}}^z \omega_{\alpha} , 
\end{equation}
which explains why $G(z, z^{\prime})$ and $G(z, \bar{z}^{\prime})$
can be chosen to have the same quadratic pieces.

\begin{figure}\label{fig3}
\centerline{\psfig{file=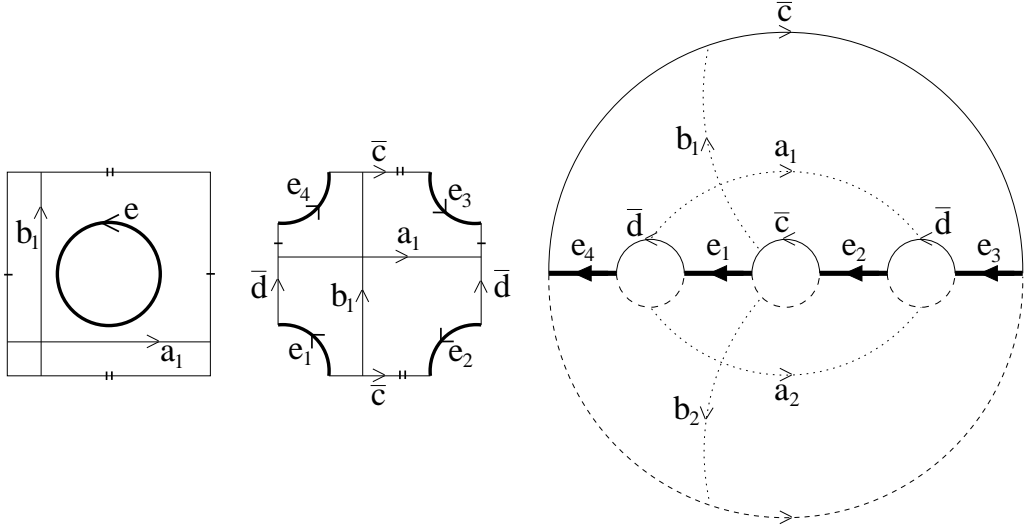,height=7cm}}
\caption{Various representations of a $(11)$ surface.  The figure
on the right side is the Schottky representation of 
a $(11)$ surface.  The overbar on a cycle denotes the fact that
the cycle is cut in half by the folding process.  The bold lines 
represent the worldsheet boundary.}
\end{figure}

The main difference between the $(03)$ surfaces and $(11)$ surfaces
is the existence of the intersecting cycles in the latter (see Fig.~3).  
In terms of the homology basis where the period matrix is
of form (\ref{period}), these cycles are written as
\begin{equation}
 {\rm a } = a_1 - b_2 ~~~ , ~~~ {\rm b} = b_1 
\end{equation}
with the intersection matrix $\CI$ given in (\ref{interm}).
We note that ${\rm a} = {\rm a_1}$ and ${\rm b } = {\rm b_1 } $ in
Fig.~3.  On top of the contractible path contribution 
to $\Omega_{\alpha}$,
we should, in general, allow the contributions from 
the integration over a cycle of the form ${\rm c} = m {\rm a} + 
n {\rm b}$ (where $mn \ne 0$), which corresponds to a nonzero cycle of 
the homology group (see Fig.~4):
\begin{equation}
 \Omega_{\alpha} = \Omega^0_{\alpha } + \Omega^t_\a
                 \equiv   \int_{\bar{z}^{\prime}}^{z} \omega_\a 
                      + \oint_{{\rm c}} \omega_\a ~ .
\label{Omega}
\end{equation}
For $\Omega^0_{\alpha}$, the integral is taken over a 
contractible path, and the second term $\Omega^t_{\alpha}$ is 
a topological number that does not change
as we locally move the positions $z$ and $z^{\prime}$.

\begin{figure}\label{scht}
\centerline{\psfig{file=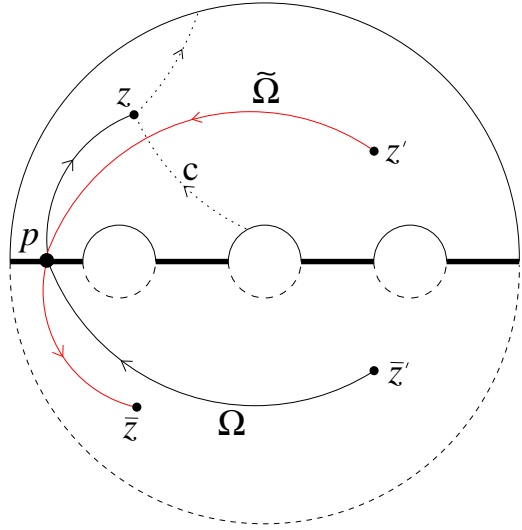,height=7cm}}
\caption{The integration path for $\Omega$ and $\tilde{\Omega}$
are shown in the Schottky representation.  The paths are 
required to pass through a reference point $P$.}
\end{figure}

Since the worldsheet propagators should be well-defined
over the whole worldsheet, we require that the $(11)$
propagators be invariant under the periodic shifts along the
${\rm p} ={\rm a}$ and ${\rm p} = {\rm b}$ cycles.  Under 
these transformations, the various objects appearing in
$(03)$ propagators in (\ref{xxprop}) and (\ref{xyprop})
change into:
\begin{eqnarray}
 \Delta \left\{ - \frac{1}{2\pi} \log \left| E (z, z') 
  \right|^2  \right\} & = & -2 \im \Omega_1 - T_{11} ~ , 
  \label{whoa1} \\
 \Delta \left\{ - \frac{1}{2\pi} \log \left| E (z, \bar{z}') 
  \right|^2  \right\} & = & -2 \im \Omega_1 - T_{11} ~ , \\
 2 \Delta \left\{ \im \Omega_\a (\tilde{T}^{-1}_B )^{\a\b} 
  \im \Omega_\b \right\} & = & 4 \frac{D_0}{D_B} \im \Omega_1
  + 2 \frac{D_0}{D_B} T_{11} ~ , \\
 2 B^2 \Delta \left\{  \re \Omega_\a (\tilde{T}^{-1}_B )^{\alpha \beta} 
 \re \Omega_\b \right\} & = & 2  \frac{B^2}{D_B} 
    ( T_{11} \re \Omega_2 - T_{12} \re \Omega_1 ) 
  + \frac{B^2}{2D_B} T_{11}   ~ , \\
 \Delta \left\{ \frac{1}{2 \pi} \log \frac{E(z, \bar{z'})}
  { \left( E ( z, \bar{z'} ) \right)^* } \right\} 
   &  = & - 2i \re \Omega_1  ~ ,  \\
 2i \Delta \left\{ \re \Omega_\a (\tilde{T}^{-1}_B )^{\alpha \beta}  
\im \Omega_\b  \right\} & = & \frac{i}{D_B} ( T_{11} \im \Omega_2
  - T_{12} \im \Omega_1 ) + 2i \frac{D_0}{D_B} \re \Omega_1  ~ ,  
\end{eqnarray}
for the shifts along the ${\rm b}$-cycle and
\begin{eqnarray}
 \Delta \left\{ - \frac{1}{2\pi} \log \left| E (z, z') 
  \right|^2  \right\} & = & 2 \im \Omega_2 - T_{22} ~ , \\
 \Delta \left\{ - \frac{1}{2\pi} \log \left| E (z, \bar{z}') 
  \right|^2  \right\} & = & 2 \im \Omega_2 - T_{22} ~ , \\
2 \Delta \left\{ \im \Omega_\a (\tilde{T}^{-1}_B )^{\a\b} 
  \im \Omega_\b \right\} & = & - 4 \frac{D_0}{D_B} \im \Omega_2
  + 2 \frac{D_0}{D_B} T_{22} ~ , \\
2 B^2 \Delta \left\{  \re \Omega_\a (\tilde{T}^{-1}_B )^{\alpha \beta} 
 \re \Omega_\b \right\} & = & 2 \frac{B^2}{D_B} 
    ( T_{22} \re \Omega_1 - T_{12} \re \Omega_2 ) 
  + \frac{B^2}{2D_B} T_{22}   ~ , \\
 \Delta \left\{ \frac{1}{2 \pi} \log \frac{E(z, \bar{z'})}
  { \left( E ( z, \bar{z'} ) \right)^* } \right\} 
   &  = & + 2i \re \Omega_2  ~ ,  \\
2 i  \Delta \left\{ \re \Omega_\a (\tilde{T}^{-1}_B )^{\alpha \beta}  
\im \Omega_\b  \right\} & = & i \frac{1}{ D_B} ( T_{22} \im \Omega_1
  - T_{12} \im \Omega_2 ) - 2 i \frac{D_0}{D_B} \re \Omega_2  ~ ,  
\label{whoa2}
\end{eqnarray}
for the shifts along the ${\rm a}$-cycle.
We use the fact that $\Omega_{\alpha}$ in (\ref{Omega}) 
transforms into $\Omega_{\alpha} + \oint_{\rm p} \omega_{\alpha}$,
the prime form remains invariant under the $a$-cycle transformation
and changes to 
\begin{equation}
 E(  \pm b_{\alpha} (z) , w | \tau ) = - \exp
 \left[ -2 \pi i ( \frac{1}{2} \tau_{\alpha \alpha} 
  \pm \int_w^z \omega_{\alpha} ) \right] E (z, w | \tau )
\end{equation} 
under the ${\rm b}$-cycle shifts as can be derived from its modular 
transformation properties.  Recalling the linear independence 
of $\re \Omega_{\alpha}$ and $\im \Omega_{\beta}$, we find 
that no combinations from (\ref{whoa1}) to (\ref{whoa2}) can 
satisfy the periodicity.

Other possible zero mode terms that we can add are the
combinations involving the off-diagonal elements of 
(\ref{useful}) proportional to the intersection matrix
$\CI$.  In particular, one can verify that
\begin{eqnarray}
- B^2 \Delta \left\{ \re \tilde{\Omega}_{\alpha} \frac{1}{D_B} 
   \CI^{\a \b}  
 \im \Omega_{\beta} +  \re \Omega_{\alpha} 
  \frac{1}{D_B} \CI^{\a \b}
 \im \tilde{\Omega}_{\beta} \right\}  \nonumber \\   =  
  - 2 \frac{B^2}{D_B}
  ( T_{11} \re \Omega^0_2 - T_{12} \re \Omega_1^0 )
 + \frac{B^2}{D_B}  \im \Omega_1^0 \equiv XX_{{\rm b}} (\Omega^0) ~ , 
    \label{t1} \\
\frac{i}{2} \Delta \left\{  \im \tilde{\Omega}_{\a} \frac{1}{D_B} 
   \CI^{\a \b} \im \Omega_{\b} - B^2 
   \re \tilde{\Omega}_{\a} \frac{1}{D_B} 
   \CI^{\a \b} \re \Omega_{\b} \right\} \nonumber \\
  =  - i \frac{1}{D_B} ( T_{11} \im \Omega^0_2 - T_{12} \im \Omega^0_1 )
  + i \frac{B^2}{2 D_B} \re \Omega_1^0  
  \equiv XY_{{\rm b}} (\Omega^0)  ~ , \label{t2}
\end{eqnarray}
for the ${\rm b}$-cycle shift and
\begin{eqnarray}
- B^2 \Delta \left\{ \re \tilde{\Omega}_{\alpha} \frac{1}{D_B} 
   \CI^{\a \b}  
 \im \Omega_{\beta} +  \re \Omega_{\alpha} 
  \frac{1}{D_B} \CI^{\a \b}
 \im \tilde{\Omega}_{\beta} \right\}  \nonumber \\   =  
  - 2 \frac{B^2}{D_B}
  ( T_{22} \re \Omega^0_1 - T_{21} \re \Omega_2^0 )
 - \frac{B^2}{D_B}  \im \Omega_2^0   
  \equiv XX_{{\rm a}} (\Omega^0) ~ , \label{t3} \\
\frac{i}{2} \Delta \left\{  \im \tilde{\Omega}_{\a} \frac{1}{D_B} 
   \CI^{\a \b} \im \Omega_{\b} - B^2 
   \re \tilde{\Omega}_{\a} \frac{1}{D_B} 
   \CI^{\a \b} \re \Omega_{\b} \right\} \nonumber \\
  =  - i \frac{1}{D_B} ( T_{22} \im \Omega^0_1 - T_{21} \im \Omega^0_2 )
  - i \frac{B^2}{2 D_B} \re \Omega_2^0  \equiv 
  XY_{{\rm a}} (\Omega^0)   ~ , \label{t4}
\end{eqnarray}
for the ${\rm a}$-cycle shift.  Here, we introduce an object
$\tilde{\Omega}_{\alpha}$ via the definition (see Fig.~4)
\begin{equation}
 \tilde{\Omega}_{\alpha} = \tilde{\Omega}^0_{\alpha } 
            + \tilde{\Omega}^t_{\alpha}              
                 \equiv \int_{z^{\prime}}^{\bar{z}} \omega_\a 
                      - \oint_{ I({\rm c})} \omega_\a ~ .
\label{Omegap}
\end{equation}
In line with the flipped sign for the topological term in comparison
to (\ref{Omega}), $\tilde{\Omega}_{\a}$ shifts to 
$\tilde{\Omega}_{\a} - \oint_{ I({\rm p}) } \omega_{\alpha}$
under a ${\rm p}$-cycle shift.  One can verify that the following
``parity" rule holds:
\begin{equation}
 \re \tilde{\Omega}^0_{\alpha} = - \re \Omega^0_{\alpha} ~~~ , ~~~ 
 \im \tilde{\Omega}^0_{\alpha} =  \im \Omega^0_{\alpha} ~~~ , ~~~
 \re \tilde{\Omega}^t_{\alpha} =  \re \Omega^t_{\alpha} ~~~ , ~~~
 \im \tilde{\Omega}^t_{\alpha} = - \im \Omega^t_{\alpha} ~ .
\label{parity}
\end{equation}
For cycles of the form ${\rm c} = m {\rm a} + n {\rm b}$ 
in (\ref{Omega}) where $m$ 
and $n$ are integers and for these cycles only,
using the explicit computation
\begin{equation}
 \oint_{{\rm a}}  \omega_{\alpha} = \pmatrix{\frac{1}{2} - i T_{12}
  \cr -i T_{22} } ~~~ , ~~~
 \oint_{{\rm b}}  \omega_{\alpha} = \pmatrix{  i T_{11}
  \cr \frac{1}{2} + i T_{12}  } ~ , 
\label{wow}
\end{equation}
it is straightforward to verify that
\begin{equation}
  XX_{{\rm b}} (\Omega^0 ) =  XX_{{\rm b}} (\Omega )  ~~~ , ~~~
  XY_{{\rm b}} (\Omega^0 ) =  XY_{{\rm b}} (\Omega ) - i m ~~~ ,
\label{t5}
\end{equation}
for the objects in (\ref{t1}) and (\ref{t2}) originating from
the ${\rm b}$-cycle shift and
\begin{equation}
  XX_{{\rm a}} (\Omega^0 ) =  XX_{{\rm a}} (\Omega )  ~~~ , ~~~
  XY_{{\rm a}} (\Omega^0 ) =  XY_{{\rm a}} (\Omega ) + i n ~~~ ,
\label{t6}
\end{equation}
for the objects in (\ref{t3}) and (\ref{t4}) coming from the 
${\rm a}$-cycle shift.   We note that the function 
$\frac{1}{2 \pi} \log \frac{E(z, \bar{z'})} 
{ \left( E ( z, \bar{z'} ) \right)^* }$ has branch cuts
since it is defined only modulo $i \IZ$.  Therefore by making 
an appropriate branch choice, we can cancel the extra 
integer terms in (\ref{t5}) and (\ref{t6}). 

Collecting the results of the analysis so far and recalling
that the effect of the constant $B$ field affects only
the zero mode parts, we can immediately write
down the following periodic worldsheet propagators for
$(11)$ surfaces:
\begin{eqnarray}
\langle X (z) X (z') \rangle \!\!\! &=& \!\!\! G(z, z') \! + \!
\frac{1 \! - \! B^2}{1 \! + \! B^2} G(z, \bar{z'})  \! + \!
\frac{2  B^2}{1 \! + \! B^2}  \re \Omega_\a 
 (\tilde{T}^{-1}_B )^{\alpha \beta} \re \Omega_\b  \nonumber \\
  &  & 
 - \frac{B^2}{1+B^2} \left( \re \tilde{\Omega}_{\alpha} \frac{1}{D_B} 
   \CI^{\a \b}  
 \im \Omega_{\beta} +  \re \Omega_{\alpha} 
  \frac{1}{D_B} \CI^{\a \b}
 \im \tilde{\Omega}_{\beta} \right)  ~ , 
\label{xx11} \\
\langle Y (z) X (z') \rangle \!\!\! &=& \!\!\!
\frac{2B}{1 \! + \! B^2}
\left( \frac{1}{2 \pi} \log \frac{E(z, \bar{z'})}
  { \left( E ( z, \bar{z'} ) \right)^* }
 + 2 i  \re \Omega_\a (\tilde{T}^{-1}_B )^{\alpha \beta}  
\im \Omega_\b ~ \right)  \nonumber \\
& & + i \frac{B}{1+B^2} \left(  \im \tilde{\Omega}_{\a} \frac{1}{D_B} 
   \CI^{\a \b} \im \Omega_{\b} - B^2 
   \re \tilde{\Omega}_{\a} \frac{1}{D_B} 
   \CI^{\a \b} \re \Omega_{\b}  \right) ~,
\label{xy11}
\end{eqnarray}
where the function $G$ is defined as
\begin{equation}
G(z,z') =  - \frac{1}{2\pi} \log \left| E (z, z') \right|^2 + 
 \im \Omega_\a (\tilde{T}^{-1}_B )^{\a\b} \im \Omega_\b ~.
\end{equation}
Using (\ref{parity}), we can rewrite
\begin{eqnarray}
 - \frac{B^2}{1+B^2} \left( \re \tilde{\Omega}_{\alpha} \frac{1}{D_B} 
   \CI^{\a \b}  
 \im \Omega_{\beta} +  \re \Omega_{\alpha} 
  \frac{1}{D_B} \CI^{\a \b}
 \im \tilde{\Omega}_{\beta} \right) \nonumber \\
 =   \frac{2 B^2}{1+B^2} \left( \re \Omega^0_{\alpha} \frac{1}{D_B} 
   \CI^{\a \b}  
 \im \Omega^t_{\beta} -  \re \Omega^t_{\alpha} 
  \frac{1}{D_B} \CI^{\a \b }
 \im \Omega^0_{\beta} \right) ~ ,
\label{b1}
\end{eqnarray}
and 
\begin{eqnarray}
 i \frac{B}{1+B^2} \left(  \im \tilde{\Omega}_{\a} \frac{1}{D_B} 
   \CI^{\a \b} \im \Omega_{\b} - B^2 
   \re \tilde{\Omega}_{\a} \frac{1}{D_B} 
   \CI^{\a \b} \re \Omega_{\b}  \right) \nonumber \\
 = i \frac{2 B}{1+B^2}  \left(  \im \Omega^0_{\a} \frac{1}{D_B} 
   \CI^{\a \b} \im \Omega^t_{\b} + B^2 
   \re \Omega^0_{\a} \frac{1}{D_B} 
   \CI^{\a \b} \re \Omega^t_{\b}  \right)    ~ .
\label{b2}
\end{eqnarray}
Noting that the part 
\begin{equation}
\frac{2  B^2}{1 \! + \! B^2}  \re \Omega_\a 
 (\tilde{T}^{-1}_B )^{\alpha \beta} \re \Omega_\b
\end{equation}
from (\ref{xx11}) and the part
\begin{equation}
  i \frac{4B}{1+B^2} \re \Omega_\a (\tilde{T}^{-1}_B )^{\alpha \beta}  
\im \Omega_\b
\end{equation}
from (\ref{xy11}) satisfy the boundary conditions \cite{klp},
we see that (\ref{b1}) and (\ref{b2}) parts also satisfy the
boundary conditions.

Given the expression for the bulk worldsheet propagators,
one can construct the boundary propagators by considering
the factorization of the string amplitudes, for example, 
as sketched in \cite{klp} for the $(03)$ surfaces.  In
this process, one should be careful to include the effects
of self-contractions.  The position of the boundary
is the line where $\re \Omega^0_{\alpha} = 0$, recalling that 
$\re \Omega^0_{\alpha} \rightarrow - \re \Omega^0_{\alpha}$
under the involution and there is a single boundary 
for the $(11)$ surfaces.  
Therefore $\re \Omega_{\alpha}$ consists of purely topological 
term $\re \Omega^t_{\alpha}$.  The covariant form of the 
boundary propagator thus obtained is as follows:\footnote{
The length dimensions of the various objects in our consideration
are $[ \theta ] = [ {\rm length} ]^2$,
$[ G_{\mu \nu} ] = [ {\rm length} ]^0$, 
$[ \tilde{T}_{\theta}^{-1} ] = [ {\rm length} ]^{-2}$,
$[ D_{\theta} ] = [ {\rm length} ]^4$ and $[ \alpha^{\prime} ] = 
[ {\rm length} ]^2$. The $\re \Omega$ and $\im\Omega$ are
dimensionless.}
\begin{eqnarray}
G^{\mu \nu}_{\rm open} (z, z^{\prime} ) & = &\alpha^{\prime} G^{\mu \nu} 
  G(z, z^{\prime}) \label{bprop} \\
 & &  +  ( \theta G \theta )^{\mu \nu}
\left( \re \Omega^t_{\alpha} ( \tilde{T}^{-1}_{\theta} )^{\alpha \beta}
 \re \Omega^t_{\beta}  + ( 2 \pi \alpha^{\prime} ) ~
  \re \Omega^t_{\alpha} ( \frac{1}{D_{\theta}} \CI )^{\alpha \beta}
 \im \Omega^0_{\beta} \right) \nonumber \\
 & & + i \theta^{\mu \nu} \Big( \frac{1}{2}  
   \epsilon ( z- z^{\prime} )
 - 2 ~ ( 2 \pi \alpha^{\prime} ) ~ \re \Omega^t_{\alpha} 
 ( \tilde{T}^{-1}_{\theta} )^{\alpha \beta} \im \Omega^{\prime}_{\beta} 
  \nonumber \\
 & &  + ( 2 \pi \alpha^{\prime} )^2 ~ \im \Omega^0_{\alpha} 
 ( \frac{1}{D_{\theta}} \CI )^{\alpha \beta} \im \Omega^t_{\beta}
  \Big) ~ , \nonumber
\end{eqnarray}
where the function $G ( z, z^{\prime} )$ is given by
\begin{equation}
 G(z , z^{\prime} ) = - \log | E( z, z^{\prime} ) |^2 
  + 2 \pi (2 \pi \alpha^{\prime} ) \im \Omega^0_{\alpha} 
( \tilde{T}^{-1}_{\theta} )^{\alpha \beta}
 \im \Omega^0_{\beta}  ~ ,
\end{equation}
where $\epsilon ( z- z^{\prime} )$ is the Heaviside step function
representing the Filk phase effect \cite{filk}.  
In (\ref{bprop}), all the integrals should be taken {\em inside} 
the $(11)$ worldsheets, while the insertion points $z$ and 
$z^{\prime}$ lie in the boundary.
The integral in $\Omega^{\prime}_\a$ is defined as 
\begin{equation}
 \Omega^{\prime}_\a = \int_P^{z} \omega^{\alpha} +
  \int_P^{z^{\prime}} \omega^{\alpha}  
\end{equation}
from a reference point $P$ on the boundary (see Fig.~4).

%%%%%%%%%%%%%%%%%%%%%%%%%%%%%%%%%%%%%%%%%%%%%%%%%%%%%%%%%%%%%%
\section{String theory amplitudes versus field theory amplitudes}
%%%%%%%%%%%%%%%%%%%%%%%%%%%%%%%%%%%%%%%%%%%%%%%%%%%%%%%%%%%%%%

Using the explicit form of the worldsheet partition
functions and the propagators now available, it is 
straightforward to compute the open string scattering
amplitudes.  In the decoupling limit, we can show that
the field theory amplitudes are reproduced from the
string theory amplitudes.  This analysis shows that the
stretched string interpretation applies to the amplitudes
involving nonplanar worldsheets.

%%%%%%%%%%%%%%%%%%%%%%%%%%%%%%%%%%%%%%%%%%%%%%%%%%%%%%%%%%%%%%
\subsection{Noncommutative field theory amplitudes}
%%%%%%%%%%%%%%%%%%%%%%%%%%%%%%%%%%%%%%%%%%%%%%%%%%%%%%%%%%%%%%

We here present various two-point 1PI Feynman amplitudes in the 
noncommutative $\phi^3$ field theory. The analyses of two-point amplitudes suffice 
the purpose of identifying the world-sheet propagator with world-line 
propagators in the field theory limits. We insert external momenta $p^\mu_1$ 
and $p^\mu_2$ into the nonplanar vacuum diagram Fig.~5(a), where three 
internal lines are labeled by the Schwinger parameters $t_a$; $a=1,2,3$; 
we employ the same momentum flow $k_a$ and the parameters $t_a$ 
in all Figures in this section. 

\begin{figure}\label{fig5}
\centerline{\psfig{file=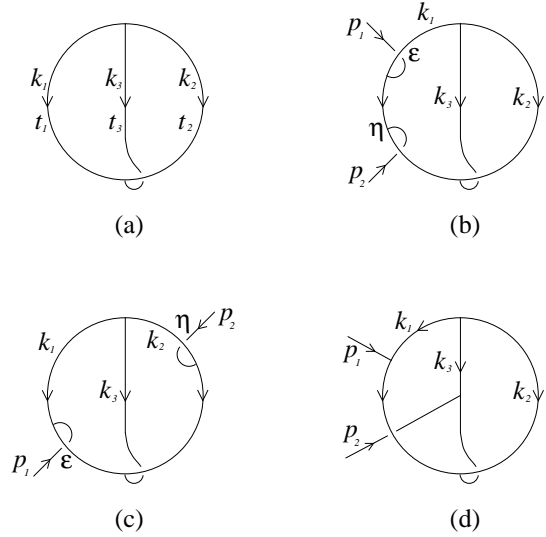,height=7cm}}
\caption{The two-loop two-point 1PI diagrams in $\phi^3$ theory.}
\end{figure}

As observed in the ordinary field theory results, it is useful 
to rearrange the Feynman amplitudes into the world-line amplitudes 
similar to string theory amplitudes. In the ordinary $\phi^3$ theory 
at two-loops, the Feynman amplitudes ($N_a$ external legs inserted on 
internal lines $t_a$) can generally be expressed as~\cite{SS}
\begin{eqnarray}
\Gamma_2^{(N_1,N_2,N_3)} 
&=& {(-g)^{N+2}\over(4\pi)^D}\cdot
\prod_{a=1}^3 \int_0^{\infty}{\rm d}t_a e^{-m^2 t_a} \cdot 
(t_1t_2+t_2t_3+t_3t_1)^{-d/2}  \nonumber \\
&\times&  \int \prod_{n=1}^{N} {\rm d}\tau_n \,   
\exp\Bigl[\,{1\over2}\sum_{a=1}^3 \sum_{j,k}^{N_a} p^{(a)}_j \cdot p^{(a)}_k 
      \, G^{\rm sym}_{aa}(\tau^{(a)}_j,\tau^{(a)}_k) \nonumber \\ 
&&   +\sum_{a=1}^3\sum_j^{N_a}\sum_k^{N_{a+1}} p^{(a)}_j \cdot p^{(a+1)}_k 
   \,G^{\rm sym}_{aa+1}(\tau^{(a)}_j,\tau^{(a+1)}_k) \,\Bigr]\ , 
\label{master}
\end{eqnarray}
where the integration regions of $\tau_n$ depend on how the Feynman diagram 
in question looks like, and the superscripts on $p_j$ and $\tau_j$ are 
just mnemonics to keep track of the internal line where those belong to. 
The world-line propagators (Green functions) $G^{\rm sym}_{ab}$; $a,b=1,2,3$, 
are essentially given by the following two functions: 
\begin{eqnarray}
G^{\rm sym}_{aa}(\tau,\tau')=|\tau-\tau'|-{t_{a+1}+t_{a+2} 
\over t_1t_2+t_2t_3+t_3t_1}(\tau-\tau')^2 \ ,\label{Gsym11}\\
G^{\rm sym}_{aa+1}(\tau,\tau')=\tau+\tau'-
{\tau^2t_{a+1}+{\tau'}^2t_{a}+(\tau+\tau')^2t_{a+2}
\over t_1t_2+t_2t_3+t_3t_1}\ . \label{Gsym13}
\end{eqnarray}

In the present cases, not only these world-line Green functions 
but also the vacuum amplitude should be modified due to the presence 
of the $\theta^{\mu\nu}$ field.  The vacuum diagram Fig.~5(a) is calculated 
by inserting a phase factor into the ordinary vacuum Feynman 
amplitude \cite{uvir,filk}: 
\begin{eqnarray}
\Gamma_2^{(a)}=g^2\int \left(\,\prod_{a=1}^3{{\rm d}^dk_a\over(2\pi)^d}
{1\over k_a^2+m^2}
\,\right) \delta^d(k_1+k_2+k_3)\,e^{i k_2\times k_3} \ ,
\end{eqnarray}
where 
\begin{eqnarray}
k\times p=k_\mu \theta^{\mu\nu} p_\nu \ ,
\end{eqnarray}
and the $\delta$-function is understood as
\begin{equation}
\delta^d(k_1+k_2+k_3)=\int{\rm d}^dy\, e^{i(k_1+k_2+k_3)}\ .
\end{equation}
Introducing parameter integrals (generally speaking the Feynman 
parametrizations) with $t_a$; $a=1,2,3$ and first performing $k_a$ 
integrals (and then $y$ integral), the above expression becomes 
the form resembling to (\ref{master}),
\begin{eqnarray}
\Gamma_2^{(a)}={g^2\over (4\pi)^d} \int^\infty_0 \!\!\! {\rm d}t_1 
\int^\infty_0 \!\!\! {\rm d}t_2 \int^\infty_0 \!\!\! {\rm d}t_3 
\;\;e^{-m^2 (t_1+t_2+t_3)}\,\, 
\mbox{det}^{1/2}\Delta_\theta \ ,  \label{whatever}
\end{eqnarray}
where $\Delta_\theta$ is the matrix given by
\begin{equation}
\Delta^{-1}_\theta = \Bigl(\,t_1t_2+t_2t_3+t_3t_1 
-\frac{\theta^2}{4} \,\Bigr)_{\mu\nu}\ .
\end{equation}
The additional $\theta^2$ factor does not appear in the cases of 
diagrams containing 
planar vacuum diagram~\cite{klp,chu,andreev}, and we expect that 
it is a purely 
topological effect inherited from string theory. Moreover, we can see a 
resemblance to the partition function (\ref{parti}) if we notice the 
reduction of ${\rm det}({\rm Im}\tau)$ to the 
factor $t_1t_2+t_2t_3+t_3t_1$~\cite{klp,RS,Lorenzo}. This will be 
more clearly explained in the next subsection. 
An important issue here is that 
we only assume the noncommutativity for spatial components (not involving 
the time component) so that the matrix $-\theta^2$ becomes positive 
definite. This corresponds to the fact that the Schwinger parameter 
$t_a$  integrals are naturally UV-regulated only when
$ - \theta^2$ is positive definite.  It is known that a field theory
with the space-space 
noncommutativity leads to perturbatively unitary results, while
the space-time noncommutativity ($ - \theta^2 < 0$) leads to the
violation of the unitarity at the field theory level \cite{Gomis}.
In the case of the light-like noncommutativity, $ - \theta^2 $ is
positive semi-definite; while this has a unitary field theory 
limit \cite{Gomis}, the stretched string interpretation 
(whose effective size is $ - \theta^2$) appears subtle.   

Now let us consider various examples of external leg insertions. 
The first example is shown in Fig.~5(b), where both external legs 
are inserted in the same internal line $t_1$. The number of ways of 
inserting a vertex are actually two, depending on how the external 
legs are attached to the internal line: going under the internal line 
or directly attached. 
According to this fact, we introduce the phase sign parameters $\epsilon$ 
and $\eta$, which take either 1 or 0. In the case of 
Fig.~5(b), these are 
assigned to be $\epsilon=\eta=1$. The corresponding Feynman amplitude 
is now calculated as 
\begin{eqnarray}
\Gamma^{(b)}_2=g^4 \int \left(\,\prod_{a=1}^3{{\rm d}^dk_a\over(2\pi)^d}
{1\over k_a^2+m^2}\,\right) 
{\delta^d(k_1+k_2+k_3)\, e^{ik_2\times k_3}\,
e^{i \epsilon k_1\times p_1} \,
e^{i \eta k_1\times p_2} \over
\Bigl((k_1+p_1)^2+m^2\Bigr)\Bigl(k_1^2 + m^2\Bigr)}\ .
\end{eqnarray}
For the external momenta $p_i$, we only assume the momentum 
conservations, not the on-shell conditions --- although the conservation 
law as such does not emerge from the momentum space representation, one can 
remember that it comes from the configuration space representation anyway. 
{}~Following the same procedures as the vacuum case, 
this amplitude can be rewritten as follows:
\begin{eqnarray}
\Gamma_2^{(b)}={g^4\over(4\pi)^d} \int^\infty_0 \!\!\! {d}t_1  
\int^\infty_0 \!\!\! {d}t_2 \int^\infty_0 \!\!\! {d}t_3  
\int^{t_1}_0\!\!\! {d}\tau_1 \int^{\tau_1}_0 \!\!\! {d}\tau_2 
\;\;\mbox{det}^{1/2}\Delta_\theta \,
\exp\left[\, p^1_\mu p^2_\nu M^{\mu\nu} \,\right]
\end{eqnarray}
with the world-line propagator (\ref{Gsym11}) modified 
\begin{equation}
M^{\mu\nu}= \left[ \vert \tau_1 -\tau_2 \vert
- \Delta_\theta\,(t_2+t_3)(\tau_1-\tau_2)^2 -(\epsilon-\eta)^2
{\theta^2\over4}\Delta_\theta\, (t_2+t_3)\, \right]^{\mu\nu}\ . 
\label{np11}
\end{equation}
Here we have omitted the mass term for simplicity of presentation: 
\begin{equation}
e^{-m^2 (t_1+t_2+t_3)} \ .
\end{equation}
It is interesting that the expression still remains symmetric in 
exchanging $t_2$ and $t_3$.  The last term in (\ref{np11}) is the
$\circ$-product term noticed in \cite{uvir}; 
We refer to the diagrams with the nonvanishing
$\circ$-product term
as nontrivial (such as $\{ \epsilon , \eta \} = \{ 0,1 \}$), and 
otherwise as trivial (when $\epsilon = \eta$).
 
In the second example (Fig.~5(c)), we insert external legs into 
the different internal lines $t_1$ and $t_2$, and the phase signs 
are assigned to be $\epsilon=\eta=1$ in the case of Fig.~5(c). 
The corresponding Feynman amplitude reads   
\begin{eqnarray}
\Gamma^{(c)}_2=g^4 \int \left(\,\prod_{a=1}^3{{\rm d}^dk_a\over(2\pi)^d}
{1\over k_a^2+m^2}\,\right) 
{\delta^d(k_1+k_2+k_3)\, e^{i(k_2+p_2)\times k_3}\,
e^{i \epsilon k_1\times p_1} \,
e^{-i \eta k_2\times p_2} \over
\Bigl((k_1+p_1)^2+m^2\Bigr)\Bigl((k_2+p_2)^2+m^2\Bigr)}\ .
\end{eqnarray}
In the same way as the first example, this can be reorganized as follows:
\begin{eqnarray}
\Gamma_2^{(c)}={g^4\over(4\pi)^d}\int^\infty_0 \!\!\! {d}t_1 
\int^\infty_0 \!\!\! {d}t_2\int^\infty_0 \!\!\! {d}t_3 
\int^{t_1}_0\!\!\! {d}\tau_1 \int^{t_2}_0\!\!\! {d}\tau_2
\;\; \mbox{det}^{1/2}\Delta_\theta \,
\exp\left[\,p^1_\mu p^2_\nu M^{\mu\nu}\, \right] \ ,
\end{eqnarray}
where the world-line propagator (\ref{Gsym13}) is modified to be 
\begin{eqnarray}
M^{\mu\nu} &=& \Bigl[\, \tau_1 + \tau_2
- \Delta_\theta\, \Bigl(\, 2t_3 \tau_1 \tau_2 + (t_2+t_3)\tau_1^2 
+ (t_1+t_3)\tau_2^2 \,\Bigr)  \nonumber\\
&-& {\theta^2\over2}\Delta_\theta\,
\Bigl(\, (\eta-1)\tau_1 +(\epsilon-1)\tau_2\,\Bigr)\nonumber\\
&-& {\theta^2\over4}\Delta_\theta\,\Bigl(\,
(\eta-1)^2t_1 + (\epsilon-1)^2t_2 + (\epsilon-\eta)^2t_3\,\Bigr) 
\Bigr]^{\mu\nu} \ . \label{pla12}
\end{eqnarray}  
We mention here that the results (\ref{np11}) and (\ref{pla12})
hold for arbitrary real numbers $\epsilon$ and $\eta$, since
we did not assume the properties $\epsilon^2 = \epsilon$
and $\eta^2 = \eta$.  

One may wonder if other diagrams such as Fig.~5(d) would give rise 
to different types of contributions.  The above two types of
expressions are general enough, however, up to the permutations. 
For example, calculating the contribution from Fig.~5(d), 
we have 
\begin{eqnarray}
\Gamma_2^{(d)}={g^4\over(4\pi)^d}\int^\infty_0 \!\!\! {d}t_1 
\int^\infty_0 \!\!\! {d}t_2\int^\infty_0 \!\!\! {d}t_3 
\int^{t_1}_0\!\!\! {d}\tau_1 \int^{t_3}_0\!\!\! {d}\tau_2
\;\;\mbox{det}^{1/2}\Delta_\theta \,
\exp\left[\,p^1_\mu p^2_\nu M^{\mu\nu}\, \right]
\end{eqnarray}
with
\begin{eqnarray}
M^{\mu\nu} &=& \Bigl[\, \tau_1 + \tau_2
- \Delta_\theta\, \Bigl(\, 2t_2 \tau_1 \tau_2 + (t_2+t_3)\tau_1^2 
+ (t_1+t_2)\tau_2^2  \,\Bigr) \nonumber \\
&-& {\theta^2\over4}\Delta_\theta\,(\, t_1+t_2 - 2\tau_1\,)\, 
\Bigr]^{\mu\nu}\ . \label{np13}
\end{eqnarray} 
This expression turns out to be a special case of (\ref{pla12}) 
for $\epsilon=1$ and $\eta=0$ with exchanging $t_2$ and $t_3$, 
or the case for $\epsilon=0$ and $\eta=1$ with exchanging $\tau_1$ and 
$\tau_2$ and the cyclic permutation $T_3\rightarrow T_2\rightarrow 
T_1(\rightarrow T_3)$.

%%%%%%%%%%%%%%%%%%%%%%%%%%%%%%%%%%%%%%%%%%%%%%%%%%%%%%%%%%%%%%
\subsection{Reduction of string theory amplitudes to field
theory amplitudes}
%%%%%%%%%%%%%%%%%%%%%%%%%%%%%%%%%%%%%%%%%%%%%%%%%%%%%%%%%%%%%%

With the worldsheet partition function and the propagator
constructed in Section 2, we can immediately write down the
string theory scattering amplitude for the two external tachyon 
insertions:
\begin{equation}
\int dy_1 dy_2 dt_1 dt_2 dt_3 
  \frac{ | W( \tau ) | }{\sqrt{ \det ( 2 \pi \alpha^{\prime} ~  
   G_{\mu \nu} ~  \im \tau
 + \frac{i}{2} \theta_{\mu \nu} \CI} ) } 
 \exp \left[ - p_{1 \mu } ~ G_{open}^{\mu \nu} ~ p_{2 \nu} 
 \right] ~ ,
\label{stringa}
\end{equation}
where $y_1$ and $y_2$ represent the two vertex insertion 
positions along the boundary, and we introduce the following 
parametrizations of the period matrix
\begin{eqnarray}
 & 2 \pi \alpha^{\prime} ~ \im \tau = \pmatrix{t_{11} & t_{12} \cr
 t_{12} & t_{22} } = \pmatrix{ t_1 + t_3 & - t_3 \cr  - t_3
   & t_2 + t_3 } ~ , & \\ & \det ~ (2 \pi \alpha^{\prime} ~ \im \tau)
 = t_1 t_2 + t_2 t_3 + t_3 t_1 . &
\end{eqnarray}
Due to the momentum conservation $p_1 + p_2 = 0$, only
the parts proportional to $G^{\mu \nu}$ in (\ref{bprop}) 
contribute to the amplitude.  Written explicitly, we
have
\begin{eqnarray}
 - p_{1 \mu } ~  G_{open}^{\mu \nu} ~ p_{2 \nu} 
& = & \alpha^{\prime} p_1 \cdot p_2 
 \log | E (y_1 , y_2 ) |^2 \label{lala} \\
& &  - p_1 \cdot p_2 ( 2 \pi \alpha^{\prime} \im \Omega^0 )_{\alpha}
 ( \tilde{T}_{\theta}^{-1} )^{\alpha \beta} 
   ( 2 \pi \alpha^{\prime} \im \Omega^0 )_{\beta}  \nonumber \\
& & - p_1 \cdot \theta^2 \cdot p_2
 \left( \re \Omega_{\alpha}^t  
  ( \tilde{T}_{\theta}^{-1} )^{\alpha \beta} \re \Omega_{\beta}^t
 +  \re \Omega_{\alpha}^t  ( \frac{1}{D_{\theta}} ) 
   \CI )^{\alpha \beta} ( 2 \pi \alpha^{\prime} \im
   \Omega^0 )_{\beta} \right) ~ , \nonumber 
\end{eqnarray}
where the dot-product and $(\theta^2 )^{\mu \nu}$ are taken
with respect to the open string metric $G_{\mu \nu}$.  

We are interested in taking the decoupling limit of Seiberg and
Witten \cite{witten}, where $\alpha^{\prime}$ goes to zero while keeping
the
open string quantities such as $G_{\mu \nu}$ and $\theta_{\mu \nu}$
fixed \cite{witten}.  In particular, we keep 
\begin{equation}
  2 \pi \alpha^{\prime} ~ \im \tau = t ~~~ {\rm and} ~~~
  2 \pi \alpha^{\prime} ~ \im \Omega^0 
\end{equation}
fixed as we take the limit.  These turn into the Schwinger
parameters and the world-line coordinates of the resulting
field theory.  The consequence of this limit, which decouples
the massive string modes, is that the
partition function part of the string theory amplitude
(\ref{stringa}) reduces to
\begin{equation}
 \frac{ | W( \tau ) | }{\sqrt{ \det ( 2 \pi \alpha^{\prime} ~  
   G_{\mu \nu} ~  \im \tau
 + \frac{i}{2} \theta_{\mu \nu} \CI} ) } 
 \rightarrow e^{ - m^2 ( t_1 + t_2 + t_3 ) } 
  \det^{1/2} \Delta_{\theta}  
\end{equation}
and the string theory quantities to
\begin{equation}
 ( \frac{1}{D_{\theta}} ) \rightarrow \Delta_{\theta}
 ~~~ , ~~~    \tilde{T}_{\theta}^{-1} \rightarrow 
 \Delta_{\theta} \pmatrix{ t_2 + t_3  & t_3 \cr
   t_3 & t_1 + t_3 } ~ ,  
\end{equation}
where $m$ is the tachyon mass and we set the open string metric
$G_{\mu \nu} = \eta_{\mu \nu} $.  A single string amplitude
can reproduce various field theory amplitudes depending
on which corner of the string moduli space one takes the 
decoupling limit.  According to \cite{RS} and
\cite{klp}, we have
\begin{eqnarray}
 2 \pi \alpha^{\prime} ~ \im \Omega^0_{12} & \rightarrow & 
    \pmatrix{ \tau_1 \cr \tau_2 }  ~ , \nonumber \\
 2 \pi \alpha^{\prime} ~ \im \Omega^0_{23} & \rightarrow & 
    \pmatrix{  \tau_3 \cr - \tau_2 - \tau_3 }   ~ , \nonumber \\
 2 \pi \alpha^{\prime} ~ \im \Omega^0_{31} & \rightarrow & 
    \pmatrix{ - \tau_1 - \tau_3 \cr \tau_3 } \label{huhu}
\end{eqnarray}  
where the indices $a$ and $b$ in $ \im \Omega^0_{b a}$ 
signify the fact that the integral is taken from the vertex in
$a$-th internal line to the vertex in $b$-th internal line
in the field theory Feynman diagrams in Fig.~5.   
Under the same circumstances, the prime form reduces to
\begin{equation}
  \alpha^{\prime} \log | E  (y_a , y_b) |^2 \rightarrow
 \tau_a + \tau_b ~ .
\end{equation}
When two insertions are made on the same internal line, for example, 
as in Fig.~5(b), the result of \cite{RS} and \cite{klp} is
\begin{equation}
  2 \pi \alpha^{\prime} ~ \im \Omega^0_{11}  \rightarrow  
    \pmatrix{ \tau_1 - \tau_2 \cr 0 } ~,
\label{puhi}
\end{equation}
and we have 
\begin{equation}
 \alpha^{\prime} \log | E  (y_1 , y_2) |^2 \rightarrow
  | \tau_1 - \tau_2 | ~ .
\end{equation}

\begin{figure}\label{boundary}
\centerline{\psfig{file=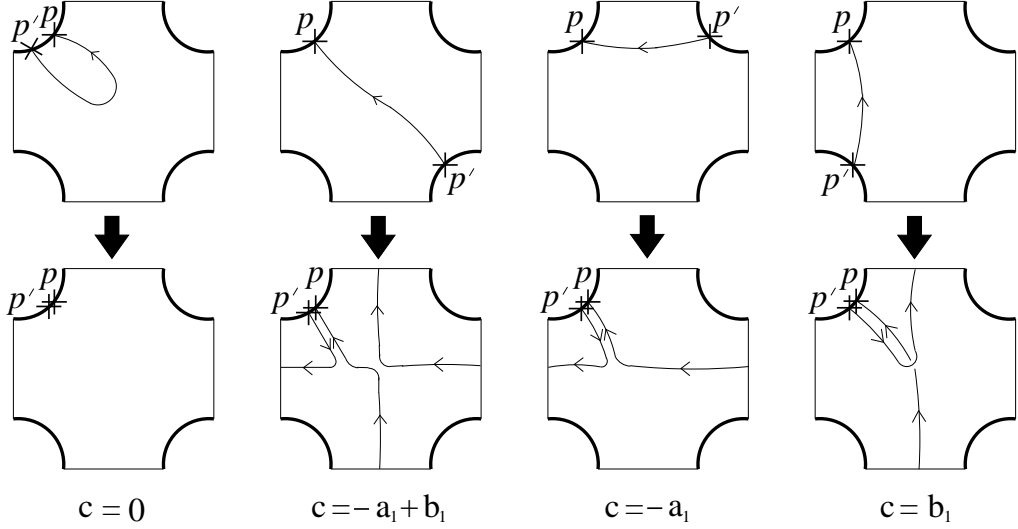,height=7cm}}
\caption{We compute the topological term that does not change
as we locally move the open string vertex by deforming
the integration path.  Three nontrivial cases are shown
along with the wound homology cycle in each case.}
\end{figure}

The computation of the topological quantity $\re \Omega^t_{1a}$
is more subtle.  As depicted in Fig.~6, we can locally move
the open string vertex $p^{\prime}$ along the boundary until 
it merges the point $p$.  Then the integration path forms a 
cycle that corresponds to one of zero (trivial in the language
of section 3.1), $- {\rm a}$, $ {\rm b }$ and $ - {\rm a} 
+ {\rm b}$ (nontrivial in the language of section 3.1) cycles.
In general, therefore, we can compute from (\ref{wow})
\begin{equation}
 \re \Omega^t_{1a} = \frac{1}{2} \pmatrix{ - \epsilon_1 \cr
  \epsilon_2 } ~ ,
\label{haha}
\end{equation}
where $\epsilon_1 = 0 ~ , ~ 1$ and $\epsilon_2 = 0 ~ , ~ 1$
corresponding to the four cases shown in Fig.~6.

To reproduce the cases of Fig.~5(c), we use (\ref{haha}) and
$ \im \Omega^0_{12}$ in (\ref{huhu}), and insert them into 
(\ref{lala}):
\begin{eqnarray}
 - p_{1 \mu } ~  G_{open}^{\mu \nu} ~ p_{2 \nu}  
   & \rightarrow & p_{1 \mu} \Bigl[\, \tau_1 + \tau_2
- \Delta_\theta\, \Bigl(\, 2t_3 \tau_1 \tau_2 + (t_2+t_3)\tau_1^2 
+ (t_1+t_3)\tau_2^2 \,\Bigr)  \nonumber\\
& & - {\theta^2\over2}\Delta_\theta\,
\Bigl(\,  -\epsilon_2 \tau_1 - \epsilon_1 \tau_2\,\Bigr)\nonumber\\
& & - {\theta^2\over4}\Delta_\theta\,\Bigl(\,
 \epsilon_2^2 t_1 + \epsilon_1^2t_2 + (\epsilon_1 
 - \epsilon_2 )^2t_3\,\Bigr) 
\Bigr]^{\mu\nu} p_{2 \nu} \ , 
\end{eqnarray}  
which shows that, upon identifying
\begin{equation}
  \epsilon_1 = 1 - \epsilon ~~~ , ~~~ \epsilon_2 = 1 - \eta ~ ,
\end{equation}
the string theory computations and the field theory computations
in (\ref{pla12}) completely agree.   To reproduce the cases
of Fig.~5(b), we use (\ref{puhi}) and (\ref{haha}) with 
$\epsilon_2 = 0$, and insert them to (\ref{lala}).  We again
see the complete agreement with the field theory result
(\ref{np11}):
\begin{equation}
 - p_{1 \mu } ~  G_{open}^{\mu \nu} ~ p_{2 \nu}  
    \rightarrow p_{1 \mu} 
  \left[ \vert \tau_1 -\tau_2 \vert
- \Delta_\theta\,(t_2+t_3)(\tau_1-\tau_2)^2 - \epsilon_1^2
{\theta^2\over4}\Delta_\theta\, (t_2+t_3)\, \right]^{\mu\nu}
  p_{2 \nu} \ ~ ,
\end{equation}
upon identifying
\begin{equation}
 \epsilon_1 = | \epsilon - \eta | ~ .
\end{equation}
In short, the general field theory results can be smoothly 
reproduced from the string theory results as one takes the 
$\alpha^{\prime} \rightarrow 0$ limit.  This fact implies
that the contribution to the loop momentum integration 
coming from the momentum region $\Delta X^{\mu} G_{\mu \nu} 
\Delta X^{\nu} < \alpha^{\prime}$ vanishes as we take the $\alpha^{\prime}
\rightarrow 0$ limit.  The stretched string interpretation
works for the field theory amplitudes built on nonplanar 
vacuum bubbles.

%%%%%%%%%%%%%%%%%%%%%%%%%%%%%%%%%%%%%%%%%%%%%%%%%%%%%%%%%%%%%%
\section{Discussions}
%%%%%%%%%%%%%%%%%%%%%%%%%%%%%%%%%%%%%%%%%%%%%%%%%%%%%%%%%%%%%%

The main finding from our analysis is that
the stretched string interpretation advocated in
\cite{liu} based on the one-loop analysis applies to the
multiloop context involving the nonplanar vacuum bubbles 
as well.  For example, the nonplanar vacuum amplitude
(\ref{whatever}) has a natural UV-regulator $\sqrt{- \theta^2}$, 
which can be interpreted as an effective stretched string 
length $\Delta X^{\mu} G_{\mu \nu} \Delta X^{\nu}$.
Combined with the results of \cite{klp} on
multiloop analysis involving the planar vacuum bubbles,
this exhausts the generic possibilities.  
Therefore, we see that the notion of stretched strings
can be naturally extended to a general multiloop context.  
In contrast to it, adding extra closed string degrees of 
freedom as suggested by \cite{uvir} appears to be
difficult to realize at the multiloop level.

One can apply the results developed in our work to
other directions; since the bulk propagator is determined
as well as the boundary propagator, it is possible to study
closed string insertions, for example, appearing in the
computation of the closed string absorption/emission 
amplitudes from noncommutative D-branes (plus closed
string loop corrections).  In the context of noncommutative
open string theory (NCOS) \cite{ncos} where the naive closed 
string coupling diverges, our approach can be directly applied to 
rigorously check its consistency against the addition
of holes to the open string worldsheet.   Furthermore,
the gluing process for the partition function 
computation sketched in Appendix can be 
straightforwardly generalized to study the cases when
some of the directions parallel to the D-branes are
compactified.  We note that the $(11)$ worldsheets produce
the field theory diagrams that show the intriguing `winding
state' behavior \cite{fisch} in the context of the thermal
field theory.  These and related issues are 
currently under investigation.

\section*{Acknowledgements}

We are grateful to Seungjoon Hyun and Sangmin Lee for helpful
discussions.  Y.~K. would like to thank Jaemo Park and 
Sangmin Lee for the collaboration at the early stage of 
this work. 

\newpage

\appendix

%%%%%%%%%%%%%%%%%%%%%%%%%%%%%%%%%%%%%%%%%%%%%%%%%%%%%%%%%%%%%%
\section*{Appendix}
%%%%%%%%%%%%%%%%%%%%%%%%%%%%%%%%%%%%%%%%%%%%%%%%%%%%%%%%%%%%%%

\section{Derivation of the (11) partition function}

The $(11)$ partition function in the presence of 
D$p$-branes can be constructed by a gluing
process starting from one-loop $(02)$ worldsheets.  In this 
appendix, for the notational simplicity, we set 
$2 \pi \alpha^{\prime} = 1$ and the open string metric 
$G_{\mu \nu} = \eta_{\mu \nu}$, where $\eta_{\mu \nu}$
is the standard Minkowskian metric.  We furthermore turn on
only the $\theta_{12} = \theta$ for the $X^1$ and $X^2$ 
(target space) spatial directions.  An annulus with the 
modulus $iT_{aa}$ is depicted in Fig.~7 where 
the two boundaries are located at $x = 0$ and $x = 1/2$.  
Along each boundary we insert a open string vertex 
and connect them.  By this
construction, a $(11)$ surface is obtained from the annulus,
a $(02)$ surface.  The (external) open string attached
to the annulus is assumed to have momentum $p_{\mu}^b$. 

\begin{figure}\label{app}
\centerline{\psfig{file=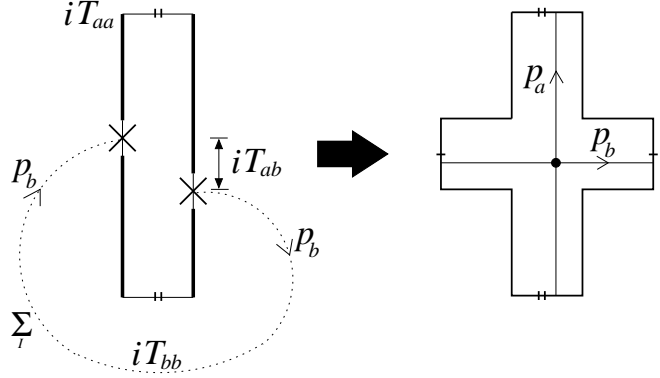,height=5cm}}
\caption{A $(20)$ surface is obtained by the nonplanar
two-point insertions along the boundaries of an annulus.}
\end{figure}

Following \cite{oneloop}, the corresponding amplitude can be
written as
\begin{equation}
\CA = \sum_I \int dp_1^b \int dp_2^b  \cdots \int dT_{ab} 
\int dT_{aa} ~ \frac{a_I}{p_1^b p_1^b + p_2^b p_2^b 
+ \cdots + M_I^2 } ~
\frac{|W_1 ( i T_{aa} ) | }{T_{aa}^{(p+1)/2}}
\label{temp1}
\end{equation}
\[ \times \exp \left[ \cdots + \frac{T_{ab}^2}{T_{aa}} 
  ( p_1^b p_1^b + p_2^b p_2 ^b )
 - \frac{\theta^2}{4T_{aa}} ( p_1^b p_1^b + p_2^b p_2 ^b ) 
\right] ~ , \]    
where $W_1$ is constructed from the one-loop eta function and
the summation over $I$ goes over the intermediate
string mass states running around the connected (external)
vertex insertions.  In (\ref{temp1}), the parameter $T_{ab}$ denotes 
the separation distance between two vertices along the imaginary
axis of the worldsheet.   Since the external vertices are
connected, the ``external" momentum $p_{\mu}^b$ is now integrated 
over.  When writing down (\ref{temp1}), we retained all the 
explicit $\theta$, $p_1^b$ and $p_2^b$ dependence, and 
the $\theta$ dependence shows up only
for the zero mode parts \cite{oneloop}.  We introduce a 
Schwinger parameter $T_{bb}$ for the ``connected leg" via
\begin{equation}
\frac{1}{p_1^b p_1^b + p_2^b p_2^b 
+ \cdots + M_I^2 } = \int dT_{bb} 
\exp \left[ - T_{bb} ( p_1^b p_1^b + p_2^b p_2^b + 
\cdots + M_I^2 ) \right] ~ ,
\label{temp2}
\end{equation}
and also introduce a ``loop momentum'' $p_{\mu}^a$ flowing
along the annulus via the
Gaussian integrals
\begin{equation}
 \sqrt{ \frac{\pi}{T_{aa}}} 
 \exp \left[ \frac{T_{ab}^2 p_1^b p_1^b - i \theta T_{ab} p_1^b p_2^b
 - \theta^2   p_2^b p_2^b / 4} {T_{aa}} \right] = 
 \int dp_1^a \exp \left[ - T_{aa} p_1^a p_1^a - 
 ( 2 T_{ab} p_1^b - i \theta p_2^b ) p_1^a \right] ~ ,   
\label{temp3}
\end{equation}
and 
\begin{equation}
\sqrt{\frac{\pi}{T_{aa}}} 
 \exp \left[ \frac{T_{ab}^2 p_2^b p_2^b + i \theta T_{ab} p_1^b p_2^b
 - \theta^2  p_1^b p_1^b /4 } {T_{aa}} \right]
=  \int dp_2^a \exp \left[ - T_{aa} p_2^a p_2^a - 
 ( 2 T_{ab} p_2^b + i \theta p_1^b ) p_2^a \right]  ~ .   
\label{temp4}
\end{equation}
Multiplying (\ref{temp2}), (\ref{temp3}) and (\ref{temp4}), 
we can rewrite (\ref{temp1}) as
\begin{equation}
\CA = \int dT_{aa} dT_{bb} dT_{ab} \sum_I  ~ \frac{|W_1 ( i T_{aa} ) | }
 {T_{aa}^{(p-1)/2}} \times \cdots
\label{temp5}
\end{equation}
\[ \times \int dp_1^a dp_1^b  dp_2^a dp_2^b 
 \exp \left[ - p_1^{\alpha} T_{\alpha \beta} p_1^{\beta}
      -   p_2^{\alpha} T_{\alpha \beta} p_2^{\beta}
 - \frac{i}{2} \theta ~ p_1^{\alpha} \CI_{\alpha \beta} p_2^{\beta}
 + \frac{i}{2} \theta ~ p_2^{\alpha} \CI_{\alpha \beta} p_1^{\beta}
\right] ~ , \]
where the imaginary part of the $(20)$ period matrix and the
intersection matrix $\CI$ are defined as 
\[ \im \tau = \pmatrix{ T_{aa} & T_{ab} \cr T_{ab} & T_{bb} } ~~~ , ~~~
    \CI = \pmatrix{ 0 & 1 \cr -1 & 0 } ~  \]
and the indices $\alpha$ and $\beta$ run over $(a,b)$. 
Performing the Gaussian integral over the $p_1^{\alpha}$ and
$p_2^{\alpha}$ yields
\begin{equation}
 \frac{1}{\sqrt{ \det ~ ( 2 \pi \alpha^{\prime} ~ G_{\mu \nu} \im \tau + 
 \frac{i}{2} \theta_{\mu \nu} \CI ) } } ~ ,
\label{temp6}
\end{equation}
where target space indices $\mu$ and $\nu$ are over $(1,2)$.
By repeating the same procedure for all the space-time
directions, we recover the partition function 
given in (\ref{parti}).
As shown in Fig.~7, the ``loop momentum" $p_{\mu}^a$ and
the external momentum $p_{\mu}^b$ intersect, thereby resulting
the matrix $\CI$ in (\ref{temp5}).  Furthermore, the original
annulus modulus $T_{aa}$, the vertex separation $T_{ab}$ and
the ``length" of the connected external leg $T_{bb}$ conspire
to form three moduli parameters of $(1,1)$ surfaces.

\end{document}